\documentclass[aps,prl,twocolumn,superscriptaddress,showpacs]{revtex4-1}
\usepackage{amsmath,amssymb,graphics,epsfig,epstopdf,color,verbatim,ulem,braket,tabularx}
\usepackage{multirow}
\usepackage[colorlinks,linkcolor=blue,citecolor=blue,urlcolor=blue,bookmarks=false]{hyperref}

\usepackage{listings}
\usepackage{cancel}
\usepackage{mathrsfs}
\usepackage{soul}
\usepackage{color}
\usepackage{url}
\usepackage{hyperref}

\makeatletter
\def\maketitle{
\@author@finish
\title@column\titleblock@produce
\suppressfloats[t]}
\makeatother

\begin{document}

\title{ Correlated electronic structure of Pb$_{10-x}$Cu$_x$(PO4)$_6$O}

\author{Changming Yue}
\email{yuecm@sustech.edu.cn}
%\affiliation{Department of Physics, University of Fribourg, 1700 Fribourg, Switzerland}
\affiliation{Department of Physics, Southern University of Science and Technology, Shenzhen 518055, China}

\author{Viktor Christiansson}
\affiliation{Department of Physics, University of Fribourg, 1700 Fribourg, Switzerland}

\author{Philipp Werner}
\email{philipp.werner@unifr.ch\\}
\affiliation{Department of Physics, University of Fribourg, 1700 Fribourg, Switzerland}

\begin{abstract}
Recently, above-room temperature superconductivity was reported in the Cu doped lead apatite Pb$_{10-x}$Cu$_x$(PO4)$_6$O, dubbed LK-99. By relaxing the structure with Cu substitution, we derive a four-band low-energy model with two 3/4 filled bands of predominantly Cu $d_{xz}$ and $d_{yz}$ character and two filled O $p_x$ and $p_y$ bands.  
This model is further downfolded to a two-band Cu-$d_{xz/yz}$ model. Using {\it ab-initio} derived interaction parameters, we perform dynamical mean field theory calculations to determine the correlated electronic structure in the normal state. These calculations yield a Mott insulator at $x=1$ and a strongly correlated non-Fermi liquid metal upon doping. The very large interaction versus bandwidth ratio $U/W\approx 30$-$50$ and the local moment paramagnetic behavior in the relevant filling regime are hard to reconcile with diamagnetism and high-temperature superconductivity. Hence, our calculations suggest that this behavior should come from a component with a different stoichiometry.     
\end{abstract}

\maketitle
 
\newpage

{\it Introduction ---}
The report of 400 K ambient-pressure superconductivity and diamagnetism in the modified lead apatite Pb$_{10-x}$Cu$_x$(PO4)$_6$O %by Lee, Kim and co-workers
 \cite{2307.12008,2307.12037} has caused considerable excitement in the physics community and beyond. Such a discovery has the potential to revolutionize many existing technologies. It is thus not surprising that experimental and theoretical groups around the world have immediately started to investigate the properties of this material, which Lee, Kim, and co-workers dubbed LK-99. Several independent groups have reported the successful synthesis of LK-99 \cite{2307.16402,2307.16802,2308.01192,2308.01516,2308.03544,2308.03110}, but the investigations of the physical properties have produced conflicting results. Some failed to find signs of superconductivity and diamagnetism \cite{2307.16402,2307.16802,2308.03544}, while others measured zero resistance below 100~K (but no Meissner effect) \cite{2308.01192} or strong diamagnetism (without measuring the resistivity) \cite{2308.01516}. In particular, Ref.~\cite{2307.16802} reported semiconducting and paramagnetic behavior. It was also pointed out that the samples might in reality be inhomogeneous and contain superconducting droplets embedded in a nonsuperconducting material \cite{2308.01723}. More work is clearly needed to clarify the many open questions, including the precise crystal structure and locations of the O and Cu atoms, the low-energy electronic structure, and the transport properties. 

Several theoretical groups have already performed electronic structure calculations using density functional theory (DFT) and DFT+$U$ \cite{2307.16040,2307.16892,2308.00676,2308.00698}. While the parent compound Pb$_{10}$(PO4)$_6$O is found to be insulating, in agreement with experiment \cite{2307.12008,2307.12037}, the substitution of a single Cu atom on a Pb(2) site (in the notation of Ref.~\cite{2307.12037}) yields two very narrow bands of Cu-$d$ character near the Fermi level. This has led to speculations about a potentially new form of flat-band high-temperature superconductivity \cite{2307.16892}. However, DFT and DFT+$U$ are clearly not sufficient for this material and electronic correlation effects beyond a standard DFT or Hartree-Fock treatment need to be taken into account. In this work, we clarify the correlated electronic structure of the normal state of Pb$_{10-x}$Cu$_x$(PO4)$_6$O by performing dynamical mean field theory (DMFT) \cite{Georges1996} calculations with {\it ab-initio} derived interaction parameters. We show that the structural relaxation in the presence of Cu dopants leads to significant changes in the lattice structure, and that the resulting low-energy model contains two bands of Cu-$d$ character near the Fermi level and two oxygen bands slightly below. This model can be further downfolded to a two-band model for the predominantly Cu $d_{xz/yz}$ bands. We calculate realistic interaction parameters using the constrained random phase approximation (cRPA) \cite{Aryasetiawan2004} and use them in DMFT calculations with a numerically exact impurity solver. The results suggest that the $x=1$ system is deep in the Mott insulating regime (interaction versus bandwidth ratio $U/W\approx 30$-$50$), while for $x>1$ and $x<1$ we find a non-Fermi liquid metal with a large local spin susceptibility.  

\begin{figure*}[t]
\includegraphics[clip,width=0.8\paperwidth,angle=0]{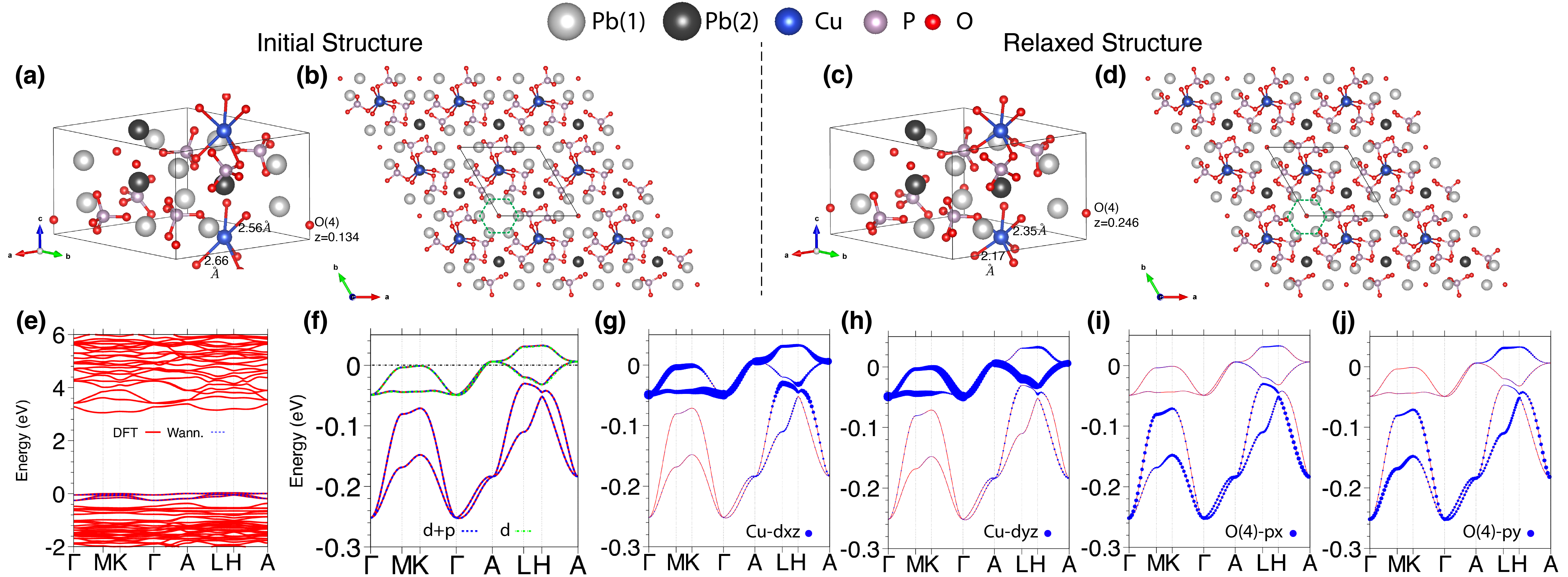}
\caption{
Crystal and band structure of Pb$_{9}$Cu(PO4)$_6$O. (a-b) [(c-d)] Crystal structure of the unrelaxed (relaxed) system. In both (a) and (c), the bond lengths between the substituted Cu atom and its nearby oxygens, as well as the fractional coordinates of O(4) along the $c$-axis, are shown. In (b) and (d), a green dashed hexagon is added to emphasize that the symmetry is lowered in the relaxed structure. (e-j) Band structure of the relaxed system, with red lines showing the DFT bands. The blue lines in (e-f) show the bands of the $d+p$ low-energy Hamiltonian obtained using the wannier90 package. The green lines in (f) further show the bands of the two-band $d$-model downfolded from the $d+p$ model using the $N$MTO method. (g-j) Orbital-resolved projected bands. 
}
\label{fig:Fig1}
\end{figure*}

{\it Low-energy model ---}
The parent compound of LK-99 is Pb$_{10}$(PO4)$_6$O$_4$, which crystallizes in the space group P6$_3$/m (176). There are 4 oxygens along the $c$-axis in Pb$_{10}$(PO4)$_6$O$_4$, whereas three of them are missing in both the undoped and doped lead apatites Pb$_{10-x}$Cu$_{x}$(PO4)$_6$O. The band structure of Pb$_{9}$Cu(PO4)$_6$O does not depend on the choice of the position (4 possible choices: $z=0.134$, 0.366, 0.634, 0.866) of the retained oxygen along the $c$-axis [dubbed O(4)] \cite{2308.00676}. 
Pb$_{9}$Cu(PO4)$_6$O contains two types of Pb atoms, referred to as Pb(1) (6 sites in the unit-cell) and Pb(2) (4 sites in the unit cell), as shown by the grey and black spheres in Fig.~\ref{fig:Fig1}. According to Refs.~\cite{2307.12008,2307.12037} the Cu atom is substituted on one of the Pb(2) sites. 
Due to the presence of inversion symmetry in Pb$_{10}$(PO4)$_6$O$_4$, there exist two non-equivalent Pb(2) positions for the Cu substitution, namely ($\frac13,\frac23,0.0035$) and ($\frac13,\frac23,0.4965$). If the structure is relaxed without relaxing the ions' positions, to preserve the space group, 
Cu should substitute the Pb(2) at the ($\frac13,\frac23,0.0035$) site and O(4) should occupy the (0,0,0.134) site, since this combination yields the minimum total energy among the enumerated choices in DFT calculations \cite{2308.00676}. This combination of Cu and O(4) is the starting point of our crystal structure relaxation, in which we relax the ionic positions but not the lattice constants. Such a relaxation is necessary because the radii of Cu and Pb are different. The substitution of Cu for Pb produces an internal strain and the atoms will readjust their positions in order to minimize the total energy and strain \cite{footnote_vca}.

In the relaxation we use the lattice constants reported in Ref.~\onlinecite{2307.12008}, which are $a=b=9.865$~\AA, $c=7.431$~\AA. The initial and the relaxed structures are shown in Fig.~\ref{fig:Fig1}(a) and \ref{fig:Fig1}(b), respectively. There are four important observations regarding the relaxed structure: (i) The symmetry is lowered to P3. In the initial structure, the Pb(1) atoms (which are not in the same plane) form a hexagonal pattern if viewed from the $c$ axis (see the green hexagon in panel (b)). However, in the relaxed structure, the Pb(1) atoms are shifted and deviate from the hexagonal pattern (see panel (d)). (ii) The O(4) is shifted along the $c$ axis from (0,0,0.134) to (0,0,0.246). (iii) The bond lengths between Cu and its adjacent oxygens shrink. (iv) The shifted O(4) atom is closer to Pb(1) (bond length reduced from 2.58~\AA \, to 2.27~\AA) but farther from Cu (distance increased from 5.76~\AA \, to 5.96~\AA). The band structure of the relaxed Pb$_{9}$Cu(PO4)$_6$O crystal is shown by the red lines in Fig.~\ref{fig:Fig1}(e). There are four bands near the Fermi-energy, which are mainly contributed by the Cu-$d_{xz}$, Cu-$d_{yz}$, O(4)-$p_x$ and O(4)-$p_y$ orbitals, as demonstrated by the band projections onto these orbitals, see panels (g-j). 

The four-band tight-binding Hamiltonian of the $d+p$-model is obtained from the maximally-localized Wannier functions constructed by wannier90 \cite{wannier90,Pizzi_2020}, and reproduces the DFT dispersion near the Fermi energy well, see blue dashed lines in panels (e) and (f). The orbital projected bands in panels (g-j) indicate that there exists some hybridization between the Cu-$d_{xz/yz}$ orbitals and the $p_{x/y}$ orbitals of O(4) \cite{footnote_wannier}.
To obtain a two-band model for Cu-$d_{xz/yz}$,  
we have to project out the O(4)-$p$ states which are almost completely occupied and presumably less relevant for the low-energy physics.
This can be achieved by performing the $N$th-order muffin-tin orbitals ($N$MTO \cite{Andersen2000}) band downfolding 
onto the Cu-$d_{xz}$ and Cu-$d_{yz}$ basis. Here we choose $N=1$ (first order).  
It turns out that the low-energy bands of predominantly Cu-$d_{xz}$ and Cu-$d_{yz}$ character can be well reproduced by this procedure, as shown by the green dot-dashed lines in Fig.~\ref{fig:Fig1}(f). We note that the resulting basis functions have a larger spatial extent
than the initial ones, but are still Wannier-like (localized, with maximal localization) and retain their specific $d$ characters \cite{Andersen2000}. 
 
In contrast to the undoped compound, the low-energy band structure of  Pb$_{9}$Cu(PO4)$_6$O -- corresponding to stacked triangular lattices -- has no particular 1D character. The nearest-neighbor hoppings between the $d$-orbitals in the $d+p$ model are $t^{(100)}_{xz,xz}=t^{(100)}_{yz,yz}=-4.65$ meV, $t^{(001)}_{xz,xz}=t^{(001)}_{yz,yz}=-3.37$ meV.
 
{\it Interaction parameters ---}
We calculate realistic interaction parameters for the two low-energy models using the constrained random phase approximation (cRPA) \cite{Aryasetiawan2004} implemented in the VASP package \cite{vasp_ref1,vasp_ref2,vasp_ref3}.
Screening from bands in an energy window from $-4.3$ to $+7.8$ eV (corresponding to roughly 100 occupied and 60 unoccupied bands, 
spin counted) are taken into account. The static ($\omega=0$) values of the density-density interactions $U$ and Hund couplings $J$ are listed in Tab.~\ref{tab_cRPA}. 
(We have also obtained qualitatively consistent, but somewhat smaller, interaction parameters using the SPEX code \cite{SPEX}.) 
As may be expected, $U$ changes appreciably between the models, while $J$ is only weakly screened. These interaction parameters are comparable to cuprates (the single-band $d$ model for La$_2$CuO$_4$ has $U_{dd}=3.65$~eV and the $d$-$p$ model $U_{dd}=7.00$~eV \cite{Werner2015}), and in particular they are much larger than the $d$ bandwidth $W\approx 0.08$~eV. For integer filling, we thus expect Mott insulating behavior. 

\begin{table}[t]
\caption{Static interaction parameters (in eV) for the $d$ and $d+p$ models obtained from cRPA.
}
\label{tab_cRPA}
\begin{tabular}{ ll|c|c } 
&&$U$&$J$\\
\hline
\hline
$d$ model &
$d$-$d$ \hspace{3mm}\mbox{}& \hspace{3mm}2.94 \hspace{3mm}\mbox{}& \hspace{3mm}0.61 \hspace{3mm}\mbox{}\\
\hline
\hline
$d+p$ model \hspace{3mm}\mbox{}& $d$-$d$ & 5.67 & 0.65\\
& $p$-$p$ & 2.66  & 0.24 \\
& $p$-$d$ & 1.36  & 0.004 \\
\end{tabular} 
\end{table}

{\it DMFT framework ---}
Focusing on the two-band model, we treat the local correlation effects in the narrow $d$ bands within the framework of DFT+DMFT  \cite{Georges1996,Kotliar2006}. This formalism 
maps the lattice problem to a self-consistent calculation of a quantum impurity problem, which is solved using continuous-time quantum Monte Carlo simulations in the hybridization expansion (CT-HYB) \cite{Werner2006,Gull2011}.
The local interactions are treated at the density-density level: $H_\text{int}=\sum_\alpha U n_{\alpha,\uparrow}n_{\alpha,\downarrow}+\sum_{\alpha\ne\beta,\sigma}[(U-2J)n_{\alpha,\sigma}n_{\beta,\bar\sigma}+(U-3J)n_{\alpha,\sigma}n_{\beta,\sigma}]$, 
where $\alpha, \beta=d_{xz},d_{yz}$ are orbital indices, $\sigma$ denotes spin and the interaction parameters are listed in Tab.~\ref{tab_cRPA}. Since the system is close to the atomic limit, the average perturbation order in the hybridization expansion is almost zero, even at temperature $T=100$~K. This creates a problem with poor Monte Carlo statistics in the standard measurement of the Green's function \cite{Werner2006}. To solve this problem, we use virtual updates \cite{Yue2020}. 

\begin{figure}[b]
\includegraphics[clip,width=3.4in,angle=0]{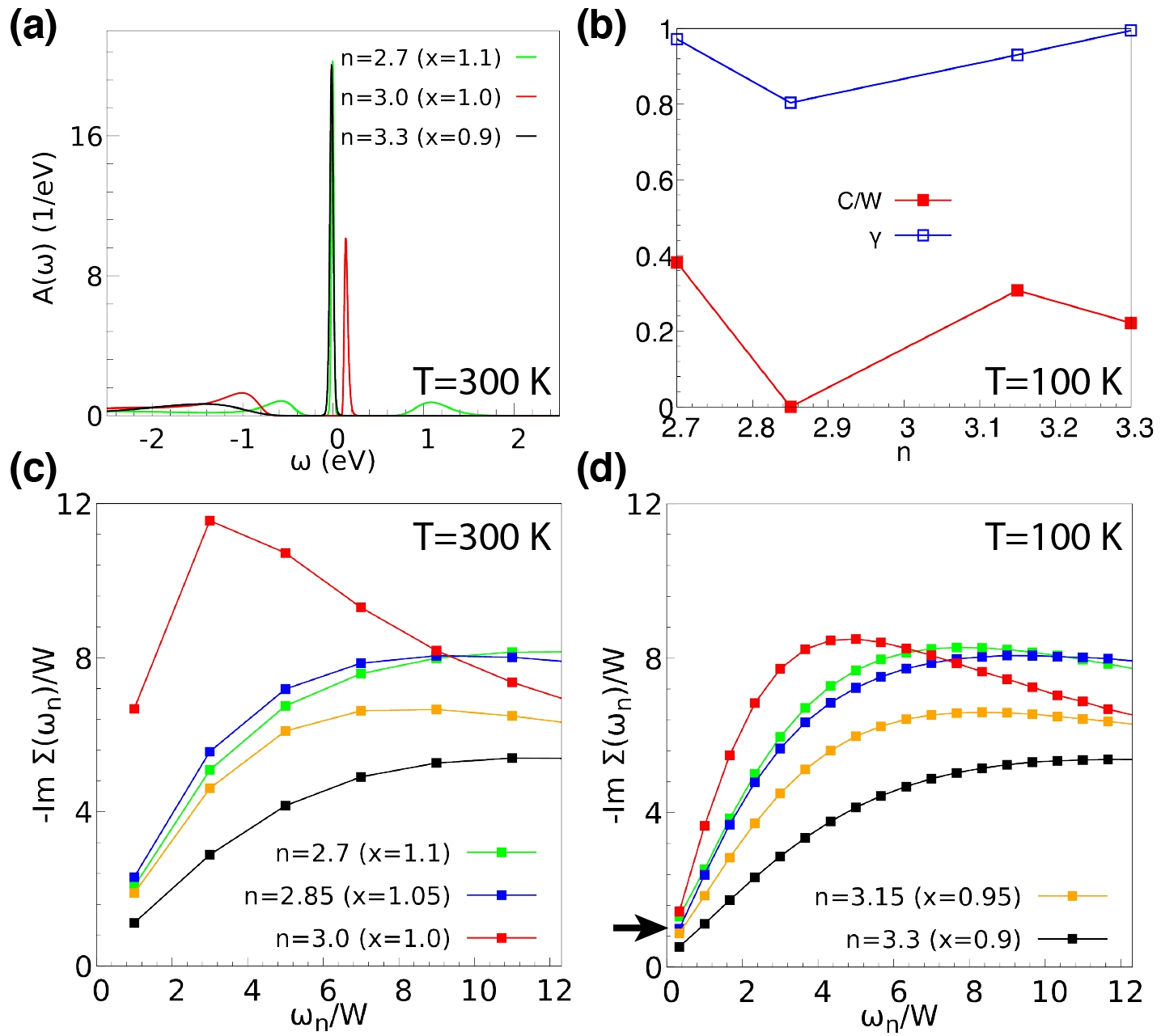}
\caption{
(a) Orbitally averaged local spectral function $A(\omega)$ at $T=100$~K. 
(b) Intercept $C$ and exponent $\gamma$, extracted from the low-frequency behavior of the Matsubara self-energy, as a function of filling.
(c,d) Imaginary part of the self-energy $-\mathrm{Im} \Sigma(i\omega_n)$ on the Matsubara frequency axis 
for $T$=300 K (c) and $T=100$ K (d).
}
\label{fig:Fig2}
\end{figure}

{\it Results ---}
Figure~\ref{fig:Fig2}(a) shows the orbitally averaged ($d_{xz}$ and $d_{yz}$ are degenerate) local spectral function obtained from the $d$ model for the indicated fillings at $T=300$~K. 
These spectra have been obtained using Maximum Entropy (MaxEnt) analytical continuation of the local imaginary-time Green's functions \cite{Jarrell1992}. Doping $x=1$ corresponds to 3/4 filling, or $n=3$ electrons in the two orbitals. At this integer filling, the system is Mott insulating, with a gap of approximately $U-3J=1.1$~eV. We notice that the spectral function is highly asymmetric, with a narrow upper Hubbard band (UHB) and a broad lower Hubbard band (LHB). This can be understood by considering the atomic limit: the UHB corresponds to the creation of quadruplons with local energy $6U-10J$ and the hopping of these quadruplons leads to a dispersion with a width comparable to the noninteracting bands ($W\approx 0.08$ eV). The LHB, on the other hand, splits into three subbands corresponding to the creation of high-spin and low-spin doublon states with local energy $U$, $U-2J$ and $U-3J$. MaxEnt cannot resolve these subbands, but the upper edge of the lower Hubbard band can be associated with the creation of high-spin doublons (local energy and gap $\approx U-3J$), and the width of the LHB is consistent with the above estimate of approximately $3J=1.8$~eV.

If we electron-dope the system ($n=3.3$, black spectrum), the chemical potential shifts into the UHB, and we obtain a quasiparticle band with width comparable to $W$, separated from the LHB by the $U-3J=1.1$~eV gap. In the hole-doped spectrum ($n=2.7$, green curve), the quasi-particle band merges with the first subband of the LHB, and we can now recognize the gap of $J=0.6$~eV to the second subband (which in the MaxEnt spectrum is merged with the third subband).  

In Fig.~\ref{fig:Fig2}(c,d) we plot the imaginary part of the orbitally-averaged self-energy $\Sigma$ as a function of Matsubara frequencies $\omega_n=(2n+1)\pi T$ for $T=300$~K (c) and $T=100$~K (d). The metallic self-energies of the higher-$T$ system look non-Fermi liquid like, while the lower-$T$ results at first sight seem to exhibit a linear behavior at low frequencies. Upon closer inspection, one notices that these almost linear curves do not extrapolate to zero as $\omega_n\rightarrow 0$, but to a finite intercept (see black arrow), indicating scattering off  frozen local moments. 
(Note that the downturn of the Matsubara self-energy for the Mott insulating $n=3$ solutions at low frequencies is a consequence of the particle-hole asymmetry of the system and does not indicate metallic behavior.) We perform an analysis analogous to Ref.~\cite{Werner2008} by fitting the low-frequency behavior of $-\text{Im}\Sigma(\omega_n)$ to the function $C+A(\omega_n)^\gamma$, with $C$, $A$ and $\gamma$ fitting parameters. For the metallic solutions, we plot the fitted values of $C$ (scattering rate) and $\gamma$ (non-Fermi liquid exponent) as a function of filling $n$ in Fig.~\ref{fig:Fig2}(b). 
%The error bars have been estimated by performing separate fits to the lowest 4 and lowest 5 data points.
The finding of a non-Fermi liquid self-energy with finite intercept (scattering rate) $C$ and $\gamma$ close to 1 both for the electron and hole-doped $n=3$ Mott insulator is consistent with the generic spin-freezing phase diagram of the two-orbital model with density-density interactions \cite{Hafermann2012}. In the large-$U$ regime, the spin-frozen bad-metal regime extends from half-filling to beyond $n=3$. 

\begin{figure}[t]
\includegraphics[clip,width=3.4in,angle=0]{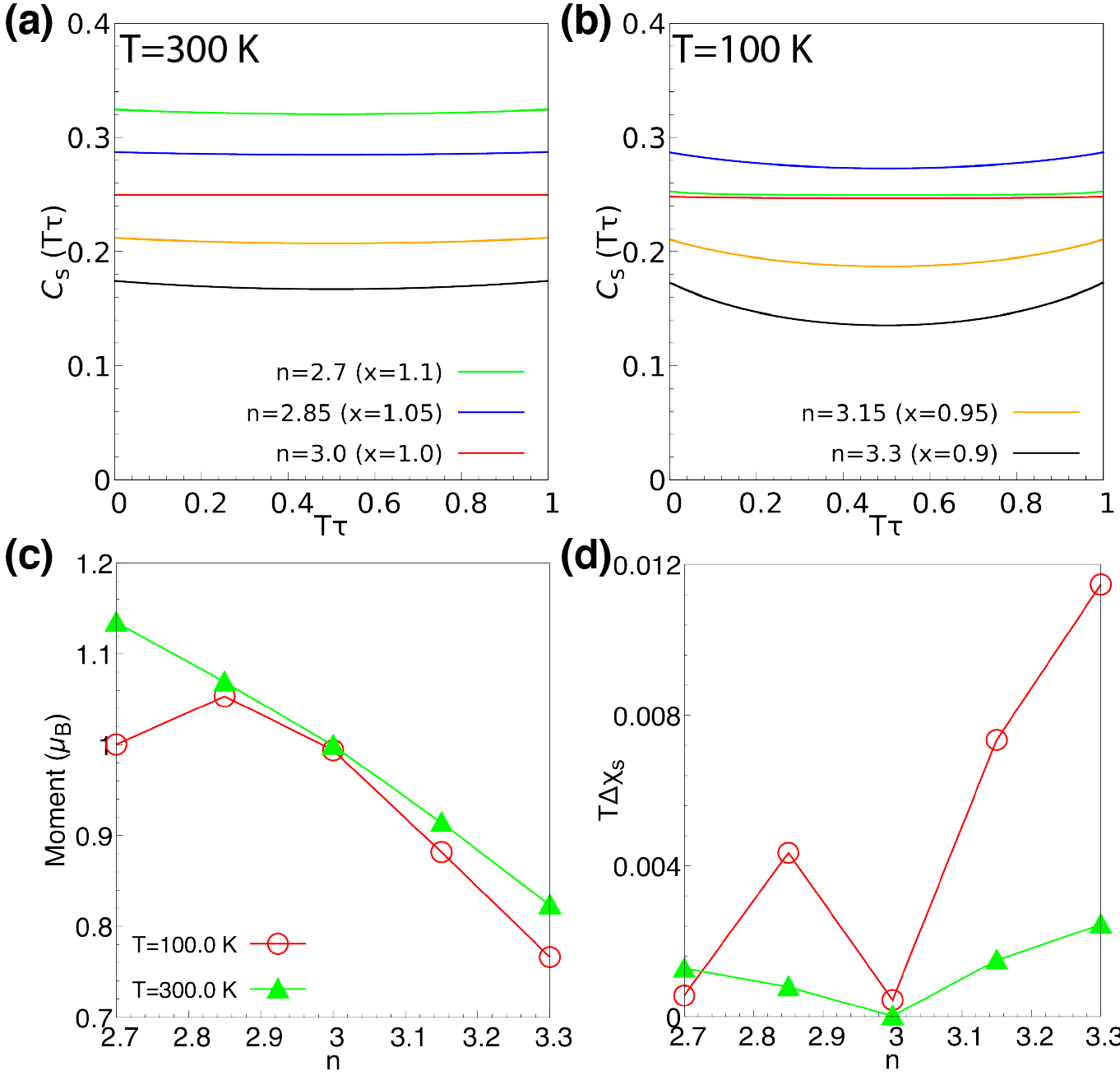}
\caption{
(a,b) Local spin-spin  correlation function $\mathcal{C}_s(\tau)=\langle S_z(\tau)S_z(0)\rangle$ of the $d$-model for the indicated fillings and $T=300$~K (a) and $100$~K (b).
(c) Estimate of the local moment 2$\sqrt{T\chi_s}$. (d) Dynamical contribution to the spin susceptibility $\Delta\chi_s=\chi_s-\frac{1}{T}\mathcal{C}_s(\frac{1}{2T})$.
}
\label{fig:Fig3}
\end{figure}

The freezing of the moments is directly seen in the local spin-spin correlation function $\mathcal{C}_s(\tau)=\langle S_z(\tau)S_z(0)\rangle$, with $S_z= \frac{1}{2} \sum_{\alpha=1}^{2}(n_{\alpha \uparrow}-n_{\alpha \downarrow})$, which is plotted in Fig.~\ref{fig:Fig3}(a,b). The flat curves imply a large local spin susceptibility $\chi_s=\int_0^{1/T} d\tau \mathcal{C}_s(\tau)$ and the presence of robust local moments (estimated in panel (c) as 2$\sqrt{T\chi_s}$). This is not only the case in the Mott insulator ($n=3$), but also in the doped systems. Since hole-doping produces high-spin moments, the freezing effect is particularly pronounced in the hole-doped metal. On the electron-doped side, the fluctuations increase, see panel (d) which plots the dynamical contributioin to the susceptibility, $\Delta\chi_s=\chi_s-\frac{1}{T}\mathcal{C}_s(\frac{1}{2T})$ \cite{Hoshino2015}. The crossover regime between the spin-frozen (or orbital-frozen) bad metal and the more conventional Fermi-liquid like metal is characterized by $\gamma\approx 0.5$ and is closely linked to unconventional superconductivity \cite{Hoshino2015,Hoshino2016,Werner2016,Hoshino2017}. In essence, the fluctuating local moments in this crossover regime provide the glue for electron pairing, while above $T_c$, they produce the bad metal behavior that is typically associated with the normal phase of unconventional superconductors. In the two-band model of LK-99, a spin-freezing crossover exists on the electron-doped side of the $n=3$ Mott insulator. However, even if we assume some doping of the Cu-$d_{xz/yz}$ bands, this material is in an unusually strongly correlated regime ($U/W\approx 30$). The most pronounced local moment fluctuations in the two-band system are found for $n\approx 2.5$ near the tip of the half-filled Mott lobe  ($U/W \approx 1$) \cite{Hoshino2016}, and also the models for cuprates \cite{Werner2016} and fulleride superconductors \cite{Hoshino2017,Yue2021} exhibit the strongest fluctuations and highest superconducting $T_c$ in this intermediate coupling regime.  

Antiferromagnetic short-range order can also lead to strange metal behavior \cite{Gunnarson2015}, and this physics is not captured by DMFT. However, in our model of LK-99, this effect should be reduced due to geometric frustration, since the Cu atoms form a triangular lattice in the plane (although the nearest distance of 7.43 \AA \, between Cu atoms is along the $c$ direction). In fact, from previous studies of two-orbital Hubbard systems, one instead expects a ferromagnetic ordering tendency for filling $n \approx 3$ in the large-$U$ regime \cite{Hoshino2016}, in agreement with the recent experimental findings in Ref.~\cite{2308.03110}.

{\it Conclusions ---}
We presented DMFT calculations for the correlated electronic structure of Pb$_{10-x}$Cu$_x$(PO4)$_6$O, focusing on the system with original O(4) position at (0,0,0.134) and with Cu substitution of the Pb(2) on the ($\frac13 , \frac23 , 0.0035$) site. We showed that the structural relaxation of this system leads to a significant shift in the O(4) position to (0,0,0.246) and derived a four-band $d+p$ model describing the physics near the Fermi level, as well as a two-band $d$ model obtained by a subsequent $N$MTO downfolding of the four-band model. In qualitative agreement with other DFT studies, we find that the two bands of predominantly Cu-$d_{xz/yz}$ character have a narrow bandwidth of approximately $0.08$ eV, which is due to the large spatial separation of 9.84~\AA \, between the Cu sites. Since the realistic interaction parameter $U$, obtained from cRPA, is about 30 times larger, it is not surprising that DMFT predicts a Mott insulator with a large gap. This is at odds with the report of high-temperature superconductivity, even if we assume some extra doping due to a somehow different stoichiometry. While unconventional superconductors (cuprates, pnictides, fullerides, \ldots) exist in proximity to a Mott state, or in a spin/orbital-freezing crossover regime which is controlled by a Mott phase \cite{Hoshino2015,Hoshino2016,Werner2016,Hoshino2017,Yue2021}, high-$T_c$ superconductivity typically appears in an intermediate correlation regime with $U/W\approx 1$ and not in the extremely correlated limit $U/W\gg 1$. 
Our LK-99 model, with filling near $n=3$ 
exhibits frozen local spin $\sim 1/2$ moments, which is not consistent with the experimentally reported diamagnetism \cite{2307.12008,2307.12037}.  
In the four-band model, depending on the choice of double counting term, we expect that a small-gap charge-transfer insulating solution could be obtained. This, together with the paramagnetism, would then look qualitatively consistent with the findings of Ref.~\cite{2307.16802}, while the physics of our models cannot be reconciled with the fascinating phenomena reported by Lee, Kim and co-workers \cite{2307.12008,2307.12037}.

The observation of flat bands in the DFT band structures has immediately raised the question of a possible manifestation of flat-band superconductivity, but it is not obvious how such a mechanism should work in the present case. At first sight, the flatness of the Cu-$d$ bands is simply a consequence of the spatial separation of the Cu atoms, and not of some nontrivial topological effect.  (The Wannier spread of the Cu orbitals of 4.08~\AA is larger than in typical $d$-electron systems, but still smaller than the lattice spacing.) 
While the quantum metric and Chern numbers of the model bands still need to be computed, the scenario discussed in Ref.~\cite{Peotta2015} is not directly applicable, since one would first have to identify a mechanism which overcomes the large repulsive interactions in the Cu-$d$ orbitals. The local moment fluctuations discussed above appear to be too weak to achieve this.

A more likely scenario is that the LK-99 samples are actually inhomogeneous, as also speculated in Refs.~\cite{2308.01723,2308.03110}. Our current model, with periodically spaced Cu dopants is certainly not very realistic. The actual material is expected to have random Cu substitutions at Pb(2) sites, according to Refs.~\cite{2307.12008,2307.12037}, and the substitution at Pb(1) sites may be energetically even more favorable \cite{2307.16892}. More importantly, one can imagine a phase separated situation with small patches accumulating higher Cu dopant concentrations. This would lead to a qualitatively different low-energy electronic structure, which may be more compatible with diamagnetism and superconductivity. More experimental investigations are needed to clarify the morphology and structure of the (superconducting) LK-99 samples, if they exist.   

{\it Acknowledgements ---}
PW thanks Cyrus Dreyer for useful discussions on the DFT band structure of LK-99, and acknowledges the Aspen Center for Physics for its hospitality during the 2023 summer program on ``bad metals."

%\clearpage
%
%\beginsupplement
%\title{Supplementary Information for\\ }
%\maketitle
%
%\begin{widetext}
%\begin{flushleft}
%\section*{DFT}
%\end{flushleft}
%\end{widetext}
 

\begin{thebibliography}{99}
\bibitem{footnote_vca} We also considered to mimic the Cu doping effect using the virtual crystal approximation (VCA), but this does not properly reproduce the charge transfer and modified bonding environment near the Cu site that we find with the explicit substitution, resulting in a very different band structure ($d$ orbitals far below the Fermi level).
\bibitem{footnote_wannier} The Wannier orbitals for O(4) have a large spread ($\sim12.9$~\AA) and are no longer centered on the original O(4) site.
\end{thebibliography}
\end{document}